# Group-velocity dispersion engineering of tantala integrated photonics

Jennifer A. Black,[1,*] Richelle Streater,[1,2] Kieran F. Lamee,[1,2] David R. Carlson,[1] Su-Peng Yu,[1,2] Scott B. Papp[1,2]

[1]*Time and Frequency Division, National Institute of Standards and Technology, Boulder, Colorado 80305, USA*
[2]*Department of Physics, University of Colorado, Boulder, CO 80309, USA*
*Corresponding author: jennifer.black@nist.gov*

**Designing integrated photonics, especially to leverage Kerr-nonlinear optics, requires accurate and precise knowledge of refractive index across the visible to infrared spectral ranges. Tantala ($Ta_2O_5$) is an emerging material platform for integrated photonics and nanophotonics that offers broadband ultralow loss, moderately high nonlinearity, and advantages for scalable and heterogeneous integration. We present refractive-index measurements on a thin film of tantala, and we explore the efficacy of this data for group-velocity dispersion (GVD) engineering with waveguide and ring-resonator devices. In particular, the observed spectral extent of supercontinuum generation in fabricated waveguides, and the wavelength dependence of free spectral range (FSR) in optical resonators provide a sensitive test of our integrated-photonics design process. Our work opens up new design possibilities with tantala, including with octave-spanning soliton microcombs.**

Photonic-integrated circuits (PICs) of waveguide-based devices offer breakthrough opportunities for basic scientific investigations, applications with light-matter interactions, and commercial devices that are widely used. In particular, waveguide PICs provide tight optical confinement and GVD engineering that is ideal for nonlinear optics. Through the Kerr effect, PIC designers may leverage efficient wavelength conversion at low input power [1], generation of chip-scale optical-frequency combs [2], and ultrabroad bandwidth supercontinuum [3]. Silicon nitride has attracted attention as a complementary metal-oxide-semiconductor (CMOS) compatible $\chi^{(3)}$ nonlinear material with nonlinear index, $n_2 = 2.4 \times 10^{-19}$ m$^2$/W [4], an order of magnitude greater than silicon dioxide [5]. Silicon-nitride devices have generated chip-scale supercontinua spanning an octave or more [6,7] as well as chip-scale octave-spanning Kerr soliton optical frequency combs [8].

Here, we explore the refractive index and the GVD design space of tantalum pentoxide ($Ta_2O_5$, hereafter tantala), a promising, CMOS-compatible $\chi^{(3)}$ nonlinear material. Indeed, tantala offers superior material properties than silicon nitride, including a higher nonlinear index $n_2 = 6.2 \times 10^{-19}$ m$^2$/W, much lower residual stress (38 MPa), a smaller thermo-optic coefficient (8.8 x 10$^{-6}$ 1/K), and a broader transparency window [9,10]. These factors make tantala an important integrated-photonics platform, and the measurements we present are critical to enable new PIC designs.

Obtaining an accurate and precise understanding of the bulk material properties is critical for future PIC design, particularly using the ion-beam-sputtered tantala films that our previous work has shown offers low-loss [11]. Previous reports of the index of refraction and absorption coefficient are available up to a wavelength of 1.8 μm [12,13], but provide inconsistent results, making it challenging to accurately design tantala PICs. To better understand tantala's optical properties and to extend this information into the mid-infrared, we have measured the refractive index and absorption coefficient. We deposited a 570 nm-thick tantala layer on 3 μm-thick thermally oxidized silicon wafer, and J. A. Woollam Co. [14] measured the layer across the spectral ranges of 190 – 1700 nm and 1.7 – 39.2 μm using RC2 and IR-VASE ellipsometry instruments, respectively. The data were fit simultaneously across both spectral regions using a combination of Sellmeier, Tauc-Lorentz, and several Gaussian dispersion functions. Figure 1(a) shows the index of refraction ($n_{TaO}$) and absorption coefficient of tantala as a function of wavelength (λ) from 200 nm to 10 μm. The results are available in Data File 1. In this spectral region, the index of refraction is predominantly larger than 2, which provides a relatively large refractive index contrast with common cladding materials (e.g. air and silicon dioxide), and the absorption coefficient is relatively low over the range 320 nm to 8 μm. Eqn. 1 presents a third-order Sellmeier equation fit

$$(1) \quad n_{TaO}^2 = 1 + \frac{0.033\lambda^2}{\lambda^2 - 0.368^2} + \frac{3.212\lambda^2}{\lambda^2 - 0.1639^2} + \frac{3.747\lambda^2}{\lambda^2 - 14.5^2},$$

describing the index of refraction data from 500 nm to 5 μm. The root-mean-square error of the fit is 0.00057.

Using Eqn. 1 as input to finite-element electromagnetic field simulations of nanophotonic waveguides, we calculate the GVD of various cross-sectional geometries of tantala PIC geometries. Figure 1(b) presents a survey of calculated GVD for straight tantala waveguide cross-sectional geometries. The simulated waveguides are a tantala core on silicon dioxide with top and side air-cladding and cross-sectional dimensions of thickness (t) and width (w). Varying the thickness and width of the waveguides makes possible anomalous GVD in various spectral regions. Figure 1(b) shows simulated waveguide thickness from 500 nm to 1750 nm, demonstrating anomalous GVD for light polarized in the chip-plane from the near-infrared into the mid-infrared, and we find that anomalous GVD is possible for straight waveguides with thicknesses of 350 nm and larger, supporting anomalous GVD at a minimum wavelength of ~770 nm.

Using the GVD information, we design and fabricate both supercontinuum-generation waveguides and ring resonators. We fabricate the devices by electron-beam lithography and plasma etching of a tantala layer deposited on a thermally oxidized silicon wafer. The top cladding is air, though an oxide cladding is possible. A deep-reactive-ion etch dices the wafer into chips and an inverse taper to the edge of the devices enhances the fundamental mode overlap for coupling optimization from a lensed fiber [15]. In this paper, we test Eqn. 1 with experiments that are highly GVD sensitive, namely the spectral extent of supercontinuum generation and the wavelength dependence of FSR in ring resonators. Our results indicate that design of tantala PICs or nonlinear devices are within fabrication tolerances.

PIC waveguides with high nonlinearity are ideal for supercontinuum generation with an input pulsed laser. For comparison, highly nonlinear optical fibers are often used to achieve such spectral broadening in silica [16], [17]. However, PIC waveguides offer much higher nonlinearity, design flexibility, and importantly geometrically tunable GVD. We design and fabricate several tantala supercontinuum waveguides on silicon dioxide with a top air cladding, a waveguide thickness of 570 nm and an optical path length of 15 mm. We design the devices to offer anomalous GVD in the 1550 nm region and two zero dispersion wavelengths for dispersive wave generation on the short- and long-wavelength side of the pump, respectively. The peak wavelength of the dispersive waves depends sensitively on the waveguide GVD and the optical pump power [16–18]. We fabricated several waveguide widths (1.1 to 1.9 μm) to provide a variation in GVD. Figure 2(a) presents a schematic of the experimental setup. We use a 100 MHz repetition rate mode-locked laser with 80 fs-duration pulses at a central wavelength of 1560 nm. We couple light into the device with a microscope objective and collect the output from the chip with a multi-mode optical fiber which directs the transmitted light to an optical spectrum analyzer.

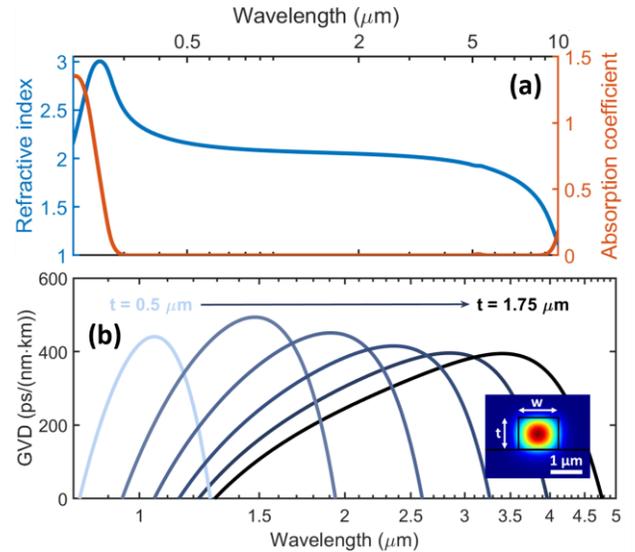

Fig. 1. (a) Refractive index ($n_{TaO}$) and absorption coefficient data for tantala from 200 nm to 10 μm on a log wavelength scale. A broad transparency window is evident from 320 nm to 8 μm. See Data File 1 for a table of values. (b) The simulated GVD for various straight tantala waveguide geometries as a function of wavelength from 800 nm to 5 μm (log scale). The waveguide thickness t varies from 0.5 to 1.75 μm in 0.25 μm steps and the waveguide width w is a constant factor 1.25 times the thickness, $w = 1.25 \times t$. The inset shows a simulation (t = 1 μm) of the fundamental transverse-electric mode with labelled cross-section dimensions (t × w).

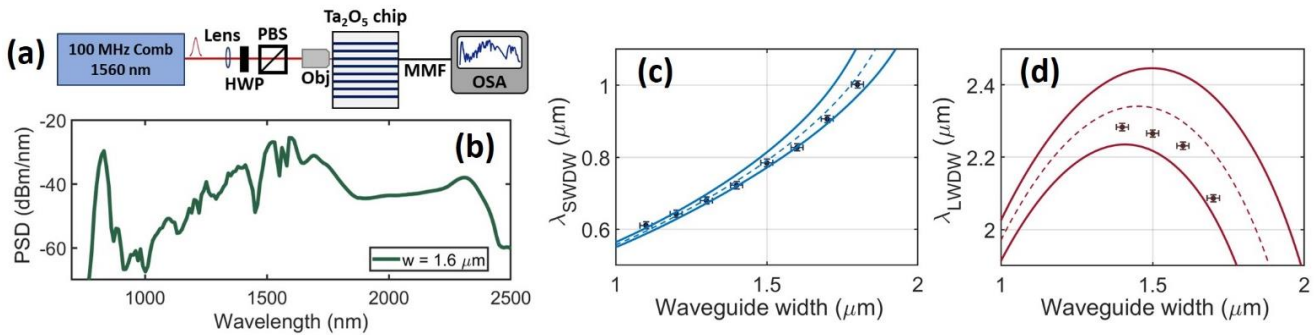

Fig. 2. (a) Experimental setup for supercontinuum measurements. A 100 MHz repetition rate optical frequency comb operating at 1560 nm (< 100 mW average power) pumps the tantala device ($Ta_2O_5$ chip). A half-wave plate (HWP) and polarizing beam splitter (PBS) prepare the optical power and polarization before the chip. A high-numerical-aperture aspheric lens (Obj) launches light onto the device, a multi-mode optical fiber (MMF) collects the transmitted light and an optical spectrum analyzer (OSA) monitors the output spectrum. (b) A sample supercontinuum spectrum (power spectral density, PSD) vs wavelength for waveguide width w = 1.6 μm. The average input pump power is 90 mW (pulse energy of 0.9 nJ). Measured zero-power dispersive wavelengths, (c) $\lambda_{SWDW}$ and (d) $\lambda_{LWDW}$ for various waveguide widths. The data (points) agree with our simulations within fabrication tolerances; dashed line is the target geometry and solid lines denote ±10 nm thickness variation.

The fabricated devices demonstrate low optical power dispersive wavelength generation. Figure 2(b) shows a sample spectrum collected from a device with waveguide width of 1.6 μm pumped with average input pump power of 90 mW (pulse energy 0.9 nJ). A fit of the maxima in the collected power spectral density away from the pump wavelength is used to find the dispersive wavelengths. A small shift in the dispersive wavelength is dependent on the pump power. To determine the zero-power dispersive wave positions of the short- and long-wavelength dispersive waves ($\lambda_{SWDW}$ and $\lambda_{LWDW}$), a power scan is performed. We find the zero-power dispersive wave positions by extrapolating a fit of the pump power dependent dispersive wavelengths. Figure 2(c) and 2(d) present $\lambda_{SWDW}$ and $\lambda_{LWDW}$ as a function of supercontinuum waveguide width. A nonlinear Schrödinger equation with GVD simulated using a semi-vectorial method is used to model $\lambda_{SWDW}$ and $\lambda_{LWDW}$. The experimental (dots) and simulated (lines) results are in good agreement. The dashed lines represent the target geometry and the solid lines signify a variation of +/- 10 nm in the waveguide thickness, demonstrating agreement within fabrication tolerances.

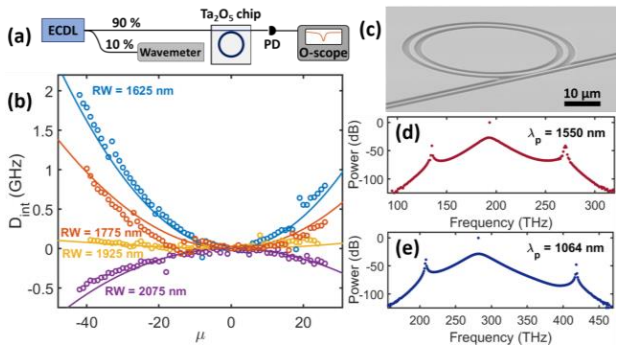

Fig. 3. (a) Schematic of the experimental setup. A widely tunable external cavity diode laser (ECDL) probes the tantala ring resonator ($Ta_2O_5$ chip). A fiber coupler sends 10 % of the optical power to a wavelength meter (wavemeter) for optical frequency measurement. A photodetector (PD) monitors the laser light transmitted through the ring resonator. Lensed fibers launch light into the device and collect light from the device. (b) The integrated dispersion ($D_{int}$, Eqn. 2) with pump wavelength near 1550 nm for ~200 GHz FSR (RR = 112.5 μm) with ring widths 1625, 1775, 1925 and 2075 nm. The GVD ranges from anomalous to normal for the various ring widths. The data (circles) and simulation (lines) are in good agreement. (c) A scanning electron microscope image of a ring resonator with 1 THz FSR (RR = 22.5 μm). LLE simulation of an octave-spanning optical frequency comb with pump wavelength (d) 1550 nm (193.4 THz) and (e) 1064 nm (281.8 THz). The ring resonator dimensions, t × w, are 570 nm × 1560 nm (415 nm × 775 nm) for pump wavelength 1550 (1064) nm and RR = 21.25 μm.

Ring resonators enhance the intensity of optical fields and provide a discrete set of propagation momenta. Ring resonators are useful for sensitive detection of bioparticles [19,20] and nonlinear processes, such as four-wave mixing and optical parametric oscillation [21,22], wavelength translation [1,23], and soliton optical-frequency-comb generation [24]. To test Eqn. 1, we fabricate tantala ring resonators and characterize the wavelength dependence of their FSR. In particular, the integrated dispersion

$$(2) \quad D_{\text{int}}(\mu) = \nu_\mu - (\nu_0 + \text{FSR} \cdot \mu)$$

is a polynomial expansion of the FSR as a function of azimuthal mode $\mu$ with respect to a reference optical frequency $\nu_0$ [25]. We calculate $D_{\text{int}}$ using a finite-element-method (FEM) mode solutions of a tantala ring resonator, including the specific tantala layer thickness, ring resonator waveguide width (RW), and ring radius (RR). We have determined that post-fabrication scanning-electron microscope images of the waveguide facets at chip edge are critical to understand the fabricated cross-sectional geometry as compared to the target design. Indeed, our current dry-etching recipe, while providing smooth and vertical sidewalls, results in a 50 nm-thick tantala residual pedestal, which we include in our FEM simulations to increase accuracy.

Figure 3 presents measurements of $D_{\text{int}}$ versus the resonator azimuthal number as defined in Eqn. 2 for several 200 GHz FSR ring resonators with a RR of 112.5 μm, measured to the center of the ring resonator waveguide. We use lensed fibers to launch light into the on-chip coupling waveguide and to collect light after the resonator. Using a wavelength-tunable, narrow-linewidth, continuous-wave laser and a wavelength meter (Fig. 3a), we record the resonant frequencies of as many fundamental transverse-electric modes as possible (Fig. 3b); we determine the resonant frequency by the minimum of resonator laser transmission. We explore the wavelength range of 1510–1620 nm. Fig. 3b presents $D_{\text{int}}$ measurements (color-coded open circles) and simulations (color-coded lines) for four specific settings of RW, namely 1625 nm (blue), 1775 nm (orange), 1925 nm (yellow), and 2075 nm (purple) that are designed to tune the GVD from anomalous to normal. In these data, degeneracies with higher-order transverse modes of the ring resonator perturb the otherwise monotonic behavior about $\mu = 0$. These results demonstrate the efficacy and accuracy of geometric GVD engineering in tantala for various applications.

A benchmark of soliton microcombs is generation of octave bandwidth spectra for f-2f self-referencing, which requires a high level of GVD engineering that has recently been accomplished with 1 THz-FSR silicon nitride ring resonators [9]. We consider octave-bandwidth designs of 1 THz-FSR tantala resonators, which we can fabricate; see the image in Fig. 3(c). We explore the theoretical design of octave-bandwidth 1 THz-FSR tantala resonators (RR = 21.25 μm) for two commonly available pump wavelengths in the infrared, 1550 nm and 1064 nm. Both have been used to demonstrate soliton optical frequency Kerr comb generation with 1-THz-FSR devices in silicon nitride [8,26]. Our design procedure uses FEM simulations to choose a waveguide geometry based on our experience of how $D_{\text{int}}$ will affect the soliton spectrum. We use the Lugiato-Lefever equation (LLE) to confirm and refine the choice of geometry [27,28]. We find the ring resonator waveguide cross-sectional dimensions for octave-spanning soliton frequency combs to be t × RW = 570 nm × 1560 nm for the 1550 nm pump wavelength and 415 nm × 737 nm for the 1064 nm pump wavelength. Figure 3(d) and 3(e) present the LLE simulated output soliton optical frequency comb spectra for pump wavelengths of 1550 and 1064 nm, respectively. We model the pump mode as critically coupled with a loaded quality factor of $1 \times 10^6$, and the on-chip optical pump power is forty times the optical parametric oscillation threshold [27]. The peak power of modes near the zero GVD crossing (dispersive waves) reside at 134.9 and 270.1 THz for the 1550 nm (193.4 THz) pump and 208.7 and 418.2 THz for the 1064 nm (281.8 THz) pump.

In conclusion, we have presented broadband refractive index and absorption coefficient data for tantala. The material provides a broad transparency window from the ultraviolet to the mid-infrared, a high index of refraction and a large nonlinearity making it an ideal candidate for fabricating linear and nonlinear integrated photonic devices. We provide an analytic third-order Sellmeier equation whose accuracy we verified by its use in designing several chip-scale nonlinear photonic devices. Anomalous GVD is possible from the near- to mid-infrared. We presented octave-spanning supercontinuum waveguides with both short and long wavelength dispersive waves and find good agreement between the experimental and simulated results. Additionally, we presented ring resonators with integrated dispersions well predicted by modelling. We show that octave-spanning soliton frequency combs are possible for pump wavelengths 1064 and 1550 nm.

**Funding.** This work was supported by NIST, DARPA A-PhI, and Advanced Research Projects Agency-Energy (ARPA-E), U.S. Department of Energy, under Award Number DE-AR0001042 program. JAB acknowledges support from the NRC Postdoctoral Fellowship.

**Acknowledgment**. We thank Pablo Acedo and Zachary Newman for technical review of the paper. We thank Ramin Lalezari for helpful information.

**Disclosures**. David Carlson is co-founder of Octave Photonics. The remaining authors do not currently have a financial interest in tantala integrated photonics.